\documentclass[12pt,preprint]{aastex}

\begin{document}
\title{Cosmic vorticity and the origin halo spins}

\author{Noam I Libeskind\altaffilmark{1}, Yehuda Hoffman\altaffilmark{2}, Matthias Steinmetz\altaffilmark{1}, Stefan Gottl\"{o}ber\altaffilmark{1}, \\ Alexander Knebe\altaffilmark{3}, \& Steffen Hess\altaffilmark{1}}
\affil{$^{1}$Leibniz-Institut f\"{u}r Astrophysik, Potsdam, An der Sternwarte 16, 14482 Potsdam, Germany}
\affil{$^{2}$Racah Institute of Physics, The Hebrew University of Jerusalem, 91904 Jerusalem, Israel}
\affil{$^{3}$Grupo de Astrofisica, Departamento de Fisica Teorica, Universidad Autonoma de Madrid, Cantoblanco E-280049, Spain}

\begin{abstract}
In the standard model of cosmology, structure emerges out of non-rotational flow and the angular momentum of collapsing halos is induced by tidal torques. The growth of halo angular momentum in the linear and quasi-linear phases is associated with a shear, curl-free, flow and it is well described within the linear framework of tidal torque theory (TTT). However, TTT is rendered irrelevant as haloes approach turn around and virialization. At that stage the flow field around halos has non-zero vorticity. Using a cosmological  simulation, we have examined the importance of the curl of the velocity field (vorticity) in determining halo spin, finding a strong alignment between the two. We have also examined the alignment of vorticity with the principle axes of the shear tensor, finding that it tends to be perpendicular to the axis along which material is collapsing fastest (${\bf e}_1$). This behavior is independent of halo masses and cosmic web environment. Our results agree with previous findings on the tendency of halo spin to be perpendicular to ${\bf e}_1$, and of the spin of (simulated) halos and (observed) galaxies to be aligned with the large-scale structure. Our results imply that angular momentum growth proceeds in two distinct phases. In the first phase angular momentum emerges out of a shear, curl-free, potential flow, as described by TTT. In the second phase, in which haloes approach virialization, the angular momentum emerges out of a vortical flow and halo spin becomes strongly aligned with the vorticity of the ambient flow field.

\end{abstract}


\section{Introduction}
\label{introduction}

Most galaxies, especially late type spirals, are observed to be rotating. Any robust theory of structure and galaxy formation needs to thus account for the origin of angular momentum.  In physics in general and hydrodynamics in particular, rotation is often associated with vorticity. Not surprisingly, the early ideas of \cite{1948ZPhy..125..290H} attempted to attribute the rotation of spiral nebulae to a primeval vorticity. The so-called theory of cosmic turbulence suggested that density perturbations, out of which structure evolves, emerged out of primordial turbulence and that galaxies got their rotation from a primordial vortical flow \citep[e.g.][cf. the excellent review of \citealt{1976RvMP...48..107J}]{1968SvA....11..907O}. These ideas were superseded by an alternative theory which has survived and emerged as the current standard model of cosmology: structure grows by the evolution of small primordial density perturbations via gravitational instability in an expanding universe \citep[e.g.][]{1974ApJ...187..425P,1978MNRAS.183..341W}. By construction the primordial flow field is a potential flow (thus curl-free) and therefore an alternative model for the origin of the galactic spin needs to be invoked. The first such ideas were developed by \cite{1934ApJ....79..460S} and \cite{1951pca..conf..195H} and then independently by \cite{1969ApJ...155..393P}, who suggested that tidal interaction between collapsing haloes may be responsible for galaxy spins. Accordingly, the induced internal velocity field in the torquing objects, is a curl-free shear flow \citep{1974MNRAS.168...73B}. It follows that the resulting angular momentum of cosmological objects is associated with a shear and not a vortical flow.

The classical picture of the origin of angular momentum was expanded by Tidal Torque Theory \citep[TTT,][]{1984ApJ...286...38W}. This seminal paper forms the base for the standard lore of the origin of angular momentum in modern cosmology. TTT provides a framework for calculating the dynamics and statistics of the growth of the angular momentum \citep[e.g.][]{1986ApJ...301...65H,1988MNRAS.232..339H,1995MNRAS.272..570S}. In fact, it is the only framework that enables an analytical analysis of the growth of the angular momentum of dark matter (DM) halos. It does so by considering the growth of the Lagrangian mass of DM halos within the linear theory of gravitational instability. Yet TTT is of limited validity when considering highly non-linear virialized halos. Using numerical simulations \citet{Porcianietal2002} examined the growth of the angular momentum of the Lagrangian mass associated with $z=0$ halos and found that TTT provides a useful description until around $\approx 60\%$ of the effective turn-around time. At $z=0$ the relative error in the predictions for the magnitude of the spin is of the order  of unity and the mean scatter in the misalignment between the simulated and the TTT predicted direction of the spin is $\approx50^\circ$. It is clear that non-linear effects weaken the applicability of TTT on the scale of virial halos. 

A close examination of TTT reveals that one of the model's most important tenets is the fact that the primordial flow field is curl-free and therefore that angular momentum emerges out of a shear flow. This principle clearly does not hold deep in the non-linear regime, where the cosmic flow field becomes vortical 
 \citep{2006MNRAS.371.1959K,2009PhRvD..80d3504P,2012MNRAS.425.2422K}.
This has prompted us to examine the possibility that the spin of halos is a result of the vorticity of the cosmic flow field on scales greater than the virial radius. This is line with recent efforts to understand the emergence of the cosmic web and the properties of DM halos in terms of the local differential properties of the cosmic flow field \citep{2012MNRAS.425.2049H,2012arXiv1210.4559L,2012MNRAS.421L.137L}

Connecting galaxy or halo properties to their environments is not a new idea, and has its roots in the seminal density-morphology relation of \cite{1980ApJ...236..351D}. A number of recent papers \citep[][among others]{Aragon-Calvoetal2007,2007MNRAS.375..184B,2012MNRAS.425.2049H,2012arXiv1210.4559L,2012MNRAS.421L.137L} have investigated the relation between halo spin and shape with the LSS in numerical simulations using a variety of different techniques to define the cosmic web. 

\cite{2012arXiv1210.4559L,2012MNRAS.421L.137L} used the velocity shear tensor to show that halo spin tends to be perpendicular to the eigenvector corresponding to the largest eigenvalue of the shear - the direction of greatest collapse (or, in voids, of weakest expansion). It is therefore natural to extended these ideas to the anti-symmetric part and examine how the direction of the curl of the velocity field is distributed relative to the eigenvectors of the shear tensor and halo spin.

\section{Methods}
\label{section:methods}

The same DM only simulation is used here as in \cite{2012MNRAS.425.2049H}. It follows the evolution of $1024^{3}$ particles in a $64 h^{-1}$Mpc periodic box employing standard WMAP5, $\Lambda$CDM cosmological parameters with a mass resolution of $\sim1.89\times 10^{7}h^{-1}M_{\odot}$ and a spatial resolution of 1$h^{-1}$kpc. The halo finder AHF \citep{knollman09} is run on the $z=0$ particle distribution. Around $\sim10^{6}$ haloes more massive than $10^{10}h^{-1}M_{\odot}$ are considered here. 

\subsection{Vorticity and shear calculation}

The velocity field is evaluated using the Clouds-In-Cell (CIC) approach calculated on a 256$^{3}$ grid giving a spatial resolution of 250$h^{-1}$kpc. The CIC gridding of the velocity field has been performed after excising the DM particles that make up virial halos. The rationale for that is to avoid 'double counting' effects when relating the properties of halos with the their ambient velocity field.  Differentiation of the velocity field is performed in Fourier space, using an FFT. The CIC velocity field is Gaussian smoothed with kernel lengths of $r_{g}=250h^{-1}$kpc, 500$h^{-1}$kpc and 1000$h^{-1}$kpc. A given smoothing length is then matched to the appropriate halo mass been, such that  $r_{g}$ roughly equals four virial radii.

The velocity field $\vec{v}$ at each point in space may by expanded as the sum of symmetric and anti-symmetric components.  The symmetric component is the shear tensor:
\begin{equation}
\Sigma_{\alpha\beta} = -\frac{1}{2H_{0}} \bigg( {\partial  v_\alpha \over \partial r_\beta}  +  {\partial  v_\beta \over \partial r_\alpha}  \bigg).
\end{equation}
where $\alpha,\beta =$ $x$, $y$, and $z$ and $H_{0}$ is the Hubble constant. Since $\Sigma_{\alpha\beta}$ is symmetric its eigenvalues ($\lambda_{1}$,~$\lambda_{2}$,~and $\lambda_{3}$) are real and its trace is equal to the divergence of the velocity, which is proportional to the overdensity in the linear regime. The number of positive (negative) eigenvalues corresponds directly to the number of axes along which matter is contracting (expanding).  If zero, one, two or three axes are simultaneously collapsing, the region may be classified as a void, sheet, filament or knot \citep[e.g. see][]{2012MNRAS.425.2049H}. These axes are just the eigenvectors of the shear tensor (${\bf e}_{1}$, ${\bf e}_{2}$ and ${\bf e}_{3}$). The anti-symmetric component of the expansion is the vorticity ${\bf \omega}$ defined as the curl of the velocity field: ${\bf \omega}=\nabla \times \bf{v}$. Vorticity generation is a strongly non-linear effect, shown by \cite{2012MNRAS.425.2422K} to appear only in third order Lagrangian perturbation theory. Initial perturbations are assumed to be effectively curl-free since in an expanding universe conservation of angular momentum implies that any vortical flow will decay linearly \citep{1995A&A...296..575B}.

Haloes are divided into three mass bins: low mass ($10^{10}- 10^{11} h^{-1}M_{\odot}$), intermediate mass ($10^{11} - 10^{12}h^{-1}M_{\odot}$) and high mass ($10^{12} - 10^{13}h^{-1}M_{\odot}$). The median halo virial radius (and standard deviation) of each mass bin are $54\pm13h^{-1}$kpc, $115\pm30h^{-1}$kpc, and $252\pm64h^{-1}$kpc, respectively. The alignment between halo spin and the vorticity is examined on scales $\gtrsim 4r_{\rm vir}$ using  the smallest (250$h^{-1}$kpc), intermediate (500$h^{-1}$kpc) and largest (1000$h^{-1}$kpc) gaussian smoothing kernel for the smallest, intermediate and largest mass bin, respectively.

\section{Results}
\label{section:results}
\subsection{Alignment of cosmic vorticity with the velocity shear. }

\begin{figure}
\begin{center}
 \includegraphics[width=40pc]{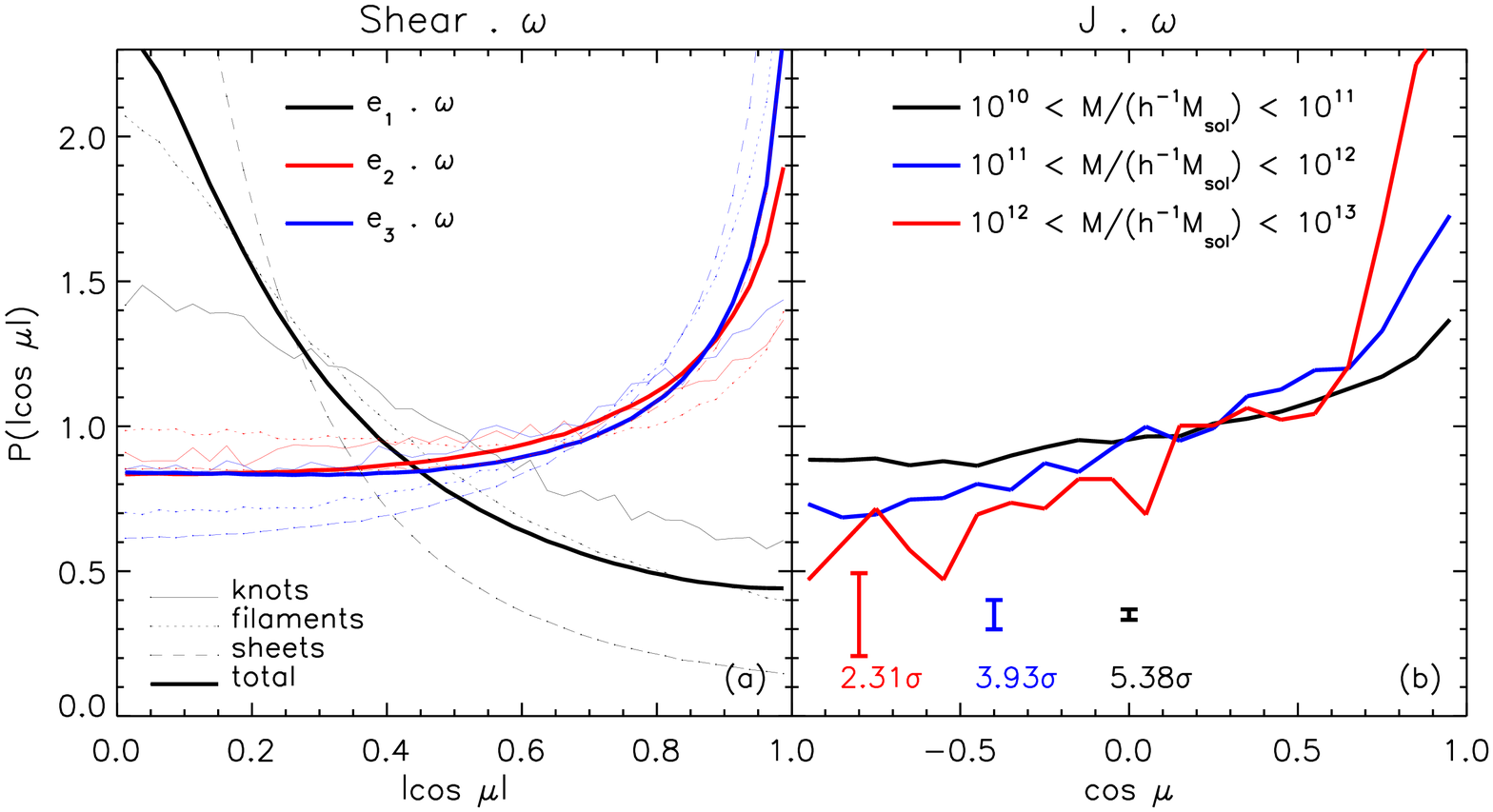}
 \caption{\textit{Left:}
 The probability distribution $P(|\cos{\mu}|)$ as a function of the angle formed, $|\cos{\mu}|$ between the vorticity and the three principle axes the shear tensor. The distribution between ${\bf \omega}$ and ${\bf e}_{1}$, ${\bf e}_{2}$, and ${\bf e}_{3}$ is shown in black, blue and red. The vorticity displays a clear perpendicular alignment with ${\bf e}_{1}$ and parallel alignment with both ${\bf e}_{2}$ and ${\bf e}_{3}$. The average offset between these distributions and a uniform one is $\sim6.5$ times the Poissonian error of a uniform distribution. \textit{Right:} The probability distribution between the vorticity and the halo spin axis $J_{\rm halo}$. Haloes are binned by mass ($10^{12}<M/h^{-1}M_{\odot}<10^{13}$ red; $10^{11}<M/h^{-1}M_{\odot}<10^{12}$ blue; and $10^{10}<M/h^{-1}M_{\odot}<10^{11}$ black). Poisson error bars for a uniform distribution with the same number of haloes in each mass bin are shown along with the average offset between the alignment and a random distribution in units of the Poisson error (i.e. the statistical significance). None of the alignments are consistent with uniform.  \label{fig:align-PD}
}
\end{center}
 \end{figure}

In this section the alignment of vorticity with the shear tensor eigenvectors is examined. The probability distribution of the angle formed between the vorticity vector and each of the three eigenvector of the shear tensor at each grid point in the simulation is shown in Fig~\ref{fig:align-PD}(a). Since the eigenvectors of the shear denote axes (not directions), the interval is confined to $|\cos\mu| \in [0,1]$. Fig~\ref{fig:align-PD}(a) clearly shows that the vorticity tends to strongly align perpendicular to ${\bf e}_{1}$ and in the ${\bf e}_{2}-{\bf e}_{3}$ plane, evidenced by the highly non uniform distributions. The apparent parallel alignment with ${\bf e}_{2}$ and ${\bf e}_{3}$ is driven by the perpendicularity with respect to ${\bf e}_{1}$. Note that there is a slight tendency for the vorticity to be more closely aligned with ${\bf e}_{3}$, than ${\bf e}_{2}$, indicating a non-uniform azimuthal distribution.

The significance of the alignment is quantified by calculating  the average offset between the measured distribution and a uniform one, in units of the Poisson error $\sigma$ per bin of a uniform distribution. If less than unity, the measured distribution is consistent with random. All three alignments in Fig.~\ref{fig:align-PD}(a) show a $\sim6.5\sigma$ statistical significance indicating they are fully inconsistent with a random distribution. The median angles between the vorticity and ${\bf e}_{1}$, ${\bf e}_{2}$ and ${\bf e}_{3}$ are 0.29, 0.60 and 0.54, respectively. \textit{The vorticity of the velocity field tends to be perpendicular to the main axis of the shear and lie in the plane of the intermediate and minor axis}. 

The alignment between the vorticity and the eigenvectors of the shear tensor is also shown for each cosmic web environment as the thin lines in Fig.~\ref{fig:align-PD}(a). The perpendicular alignment between ${\bf \omega}$ and ${\bf e}_{1}$ decreases in strength from sheets, to filaments, to knots. This is likely due to the increased erratic behavior of these vectors deep in non-linear regimes. Regardless of which web element is examined, ${\bf \omega}$ tends to cluster more towards ${\bf e}_{3}$ than ${\bf e}_{2}$ and always away from $\hat{\bf e_{1}}$.
 
Given the dominance of the shear tensor in shaping the local dynamics it is only natural to study the orientation of the vorticity with respect to its three eigenvectors. Since the eigenvectors are non-directional lines (axes) the coordinate system they define does not span the full three-dimensional volume, but just the positive octant of the cartesian grid. Each vorticity vector can thus be normalized to unity, and its direction projected onto a unit sphere defined by the eigenvectors. 

\begin{figure}
 \begin{center}
 \includegraphics[width=35pc]{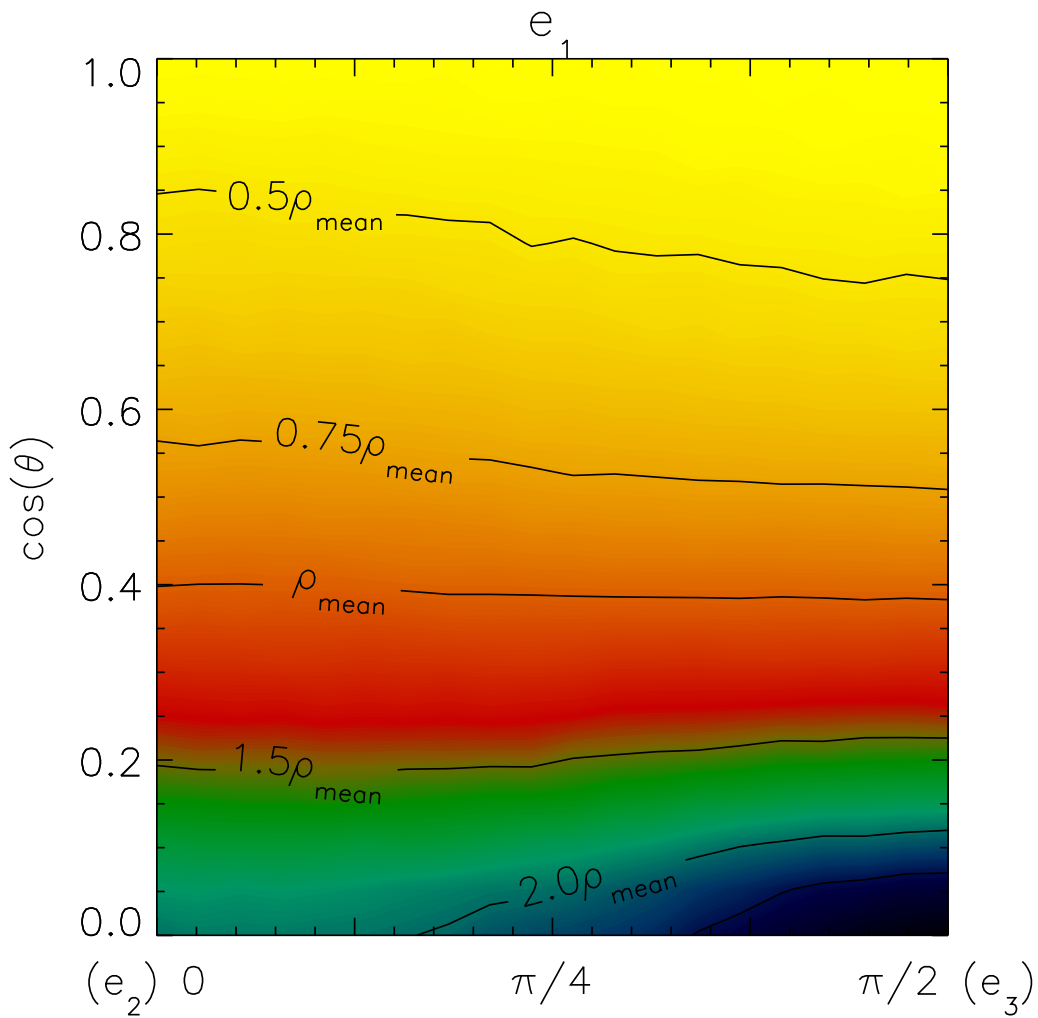}
\end{center}
 \caption{The orthonormal eigensystem of the shear can be used to define a coordinate system. By normalizing the vorticity vectors to unity, their orientation with respect to this coordinate system can be shown to be non-uniform.  One eighth of the unit sphere is shown in an equal area projection here, with contours and colors denoting the (surface) density of vorticities. ${\bf e}_{1}$ points towards the north pole (i.e. $\cos(\theta)=1$) while ${\bf e}_{2}$ and ${\bf e}_{3}$ are longitudes of $0$ and $\pi/2$, respectively. A uniform distribution is represented by a monochromatic projection. \label{fig:projection}
}
 \end{figure}

The distributions of ${\bf \omega}$ on (one eighth of) the unit sphere can be seen in an equal area projection in Fig.~\ref{fig:projection}. The perpendicularity with respect to ${\bf e}_{1}$ is clearly visible as the vorticity tends to inhabit regions close to the ${\bf e}_{2}$-${\bf e}_{3}$ plane (the ``equator''). As exemplified by the stronger parallel alignment between ${\bf \omega}$ and  ${\bf e}_{3}$ in Fig.~\ref{fig:align-PD}(a), at low latitude (i.e. for ${\bf \omega}$'s that are far from ${\bf e}_{1}$), the vorticity is closer to ${\bf e}_{3}$ than to ${\bf e}_{2}$. At higher latitudes, the locus of vorticity vectors is more uniform moving closer to ${\bf e}_{2}$ at the highest latitudes.

\begin{figure}
 \begin{center}
 \includegraphics[width=35pc]{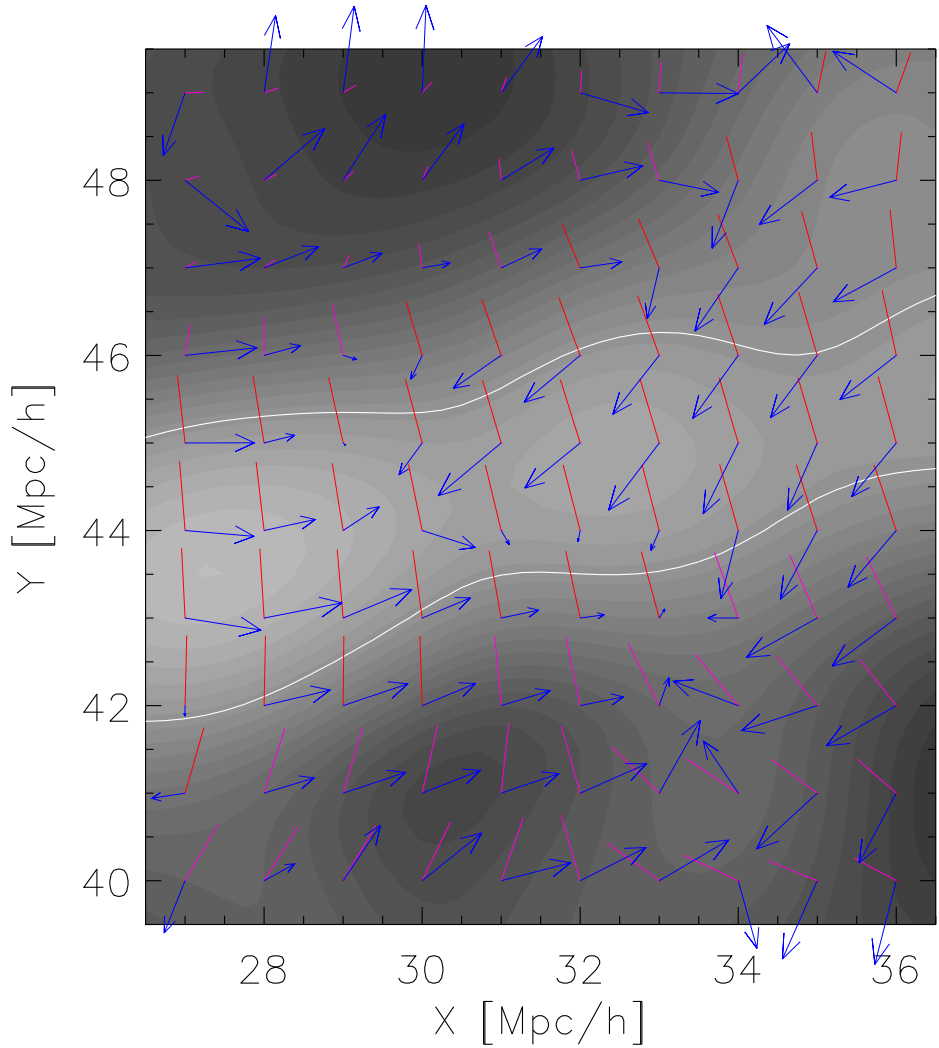}
  \caption{A 250$h^{-1}$kpc slice through the simulation showing the unit vorticity vector (${\bf \omega}$) (blue arrows) and the principle axis (${\bf e}_{1}$) of the shear tensor. Regions where ${\bf e}_{1}$ corresponds to collapse (expansion) are shown in red (magenta). Since ${\bf e}_{1}$ is an axis and not a vector, arrowheads are not plotted. The density is shown as the grayscale and the white contour denotes overdensity with respect to the mean. The length of the ${\bf e}_{1}$ and ${\bf \omega}$ vectors are normalized to unity such that a short vector implies perpendicularity with respect to the plane.}
  \label{fig:slice}
\end{center}
 \end{figure}

In Fig.~\ref{fig:slice} we present the vorticity and shear field around a generic overdensity in a $10\times10h^{-1}$Mpc region. In overdense regions (where the vorticity is well defined) ${\bf e}_{1}$ tends to point in the same direction across many Mpc, indicating that the axis of greatest collapse is correlated on these scales. In the same region, the perpendicularity of the vorticity and ${\bf e}_{1}$ is clearly visible.

\subsection{Alignment of cosmic vorticity with halo spins. }

Since \cite{2012arXiv1210.4559L} showed that halo spin axes tend to align with the principle axes of the shear tensor, and in the preceding section a correlation between the shear eigenvectors and the vorticity was shown, the alignment of vorticity and halo spin is now examined. Such an alignment would be indicative of a rotational flow, rather than the classical picture that haloes form out of a potential, irrotational flow. In Fig.~\ref{fig:align-PD}(b) we show the alignment of halo spin with the vorticity for three halo mass bins. Note that in this case the angle is between two vectors and as such allowed to range from [-1,1]. 

Fig.~\ref{fig:align-PD}(b) shows a clear correlation of halo spin with vorticity for all mass bins. The strength of this alignment is mass dependent, with larger halo spin being more tightly aligned with the vorticity: the smallest, intermediate and highest mass bins have medians of $\cos\mu=$0.11, 0.22, and 0.37. For the smallest mass bin, the haloes have the weakest alignment, yet display the strongest signal ($\sim 5.3\sigma$) statistically speaking, since this mass bin contains the largest number of haloes. The highest mass bin is inconsistent with a random distribution at the $\sim$98\% confidence level and shows the strongest alignment. \textit{Halo spin is strongly correlated with the large scale vorticity of the velocity field}.

 \section{Conclusions and discussion}

Using a high resolution cosmological simulation the relationship between halo spin and the relevant directions that characterize the large scale structure (LSS) is studied. The velocity field has been evaluated on a smoothed, clouds-in-cells grid. The local differential behavior of the velocity field is then accounted for, at each grid cell, by the symmetric and antisymmetric components. The directions of the vorticity, the eigenvectors of the shear tensor (which represent directions of collapse or expansion) and halo angular momentum have been compared. The main findings are threefold: 
\begin{enumerate}
\item Halo spin is strongly correlated with the direction of the local vorticity on scales $\gtrsim4r_{\rm vir}$. The strength of the correlation is mass dependent, with the most massive haloes being the most aligned with the vorticity.
\item The direction of the vorticity tends to be perpendicular to the shear eigenvector associated with the largest eigenvalue ($\hat{\bf{e}}_1$). In all web environments except voids (namely knots, filaments and sheets) $\hat{\bf{e}}_1$ corresponds to the axis of fastest collapse. In voids $\hat{\bf{e}}_1$ corresponds to the axis of slowest expansion.
\item The strength of both alignments depends only weakly on the web classification and the general trends are found throughout the simulated volume.
\end{enumerate}

A key element of the standard model of cosmology is that structure emerges out of rotation-free Gaussian primordial perturbations into a homogenous and isotropic universe. It follows that the early dynamical phase of halo growth is characterized by being a shear, curl-free, (potential) flow. Tidal Torque Theory \cite[TTT,][]{1984ApJ...286...38W} provides the theoretical framework to account for how the angular momentum of a halo's Lagrangian mass grows at this early stage. However, as the Lagrangian mass of a halo decouples from the Hubble expansion, the flow field around the halo becomes more non-linear and more vortical \citep[see][]{2012MNRAS.425.2422K}. \cite{Porcianietal2002} showed that at about half the halo's turn around time, the growth of the halo's angular momentum no longer follows TTT's predictions. By analyzing a high resolution DM-only cosmological simulation we have shown here that at the present epoch the spin of simulated halos is strongly aligned with the vorticity of the large scale velocity field. This suggests that the acquisition of a halo's spin proceeds in two phases, before and after roughly half of the halo's turn around time. In the first phase the halo is in its linear and quasi-linear regime and the angular momentum grows out of an irrotational shear flow. In the second phase as the halo proceeds towards a virial equilibrium, it condenses out of a rotational, vortical flow. At that final stage the spin of a halo reflects the vorticity of the flow field of its immediate neighborhood.

The orientation of galaxies and clusters with respect to the LSS has been shown to be an important variable in the process of galaxy formation and has repercussions on for example, the correlation function \citep[e.g. see][]{2012MNRAS.424.2954V}. Our findings of a strong alignment of halo spin with the vorticity, and the tendency of the vorticity to be orthogonal to $\hat{\bf{e}}_1$ regardless of the web classification, can thus be directly compared with observational studies.

For example, \cite{NavarroAbadiSteinmetz04} reported that nearby disc galaxies (including the Milky Way) tend to spin within the super-galactic plane. On greater scales, \cite{1982A&A...107..338B} showed that the orientations of clusters of galaxies are anisotropic with respect to the cluster distribution: prolate clusters are roughly parallel. Using the 2MASS survey \cite{LeeErdogdu2007} reconstructed the observed shear field finding an alignment between galaxy spins and the intermediate axis. Alignments have been examined in the SDSS by \cite{2011MNRAS.414.2029P} among others: \cite{2012arXiv1207.0068T} found that the spin of spiral galaxies is aligned with the filament axis (i.e. perpendicular to $\hat{\bf{e}}_1$). The orientation of spin with respect to voids is controversial: \cite{2006ApJ...640L.111T} reported that disc galaxies located in shells surrounding voids, rotate preferentially in the shell. This was disputed by \cite{2009JCAP...06..009S} who found a distribution consistent with random. \cite{2012ApJ...744...82V} find that within specifically chosen shells, discs are aligned perpendicular to the void's radial direction. In our web classification scheme voids are bounded by sheets, for which the plane of the sheet is defined by $\hat{\bf{e}}_1$. Since the vorticity tends to lie within the plane of the sheet, it follows that  halo spins are embedded within sheets  - an alignment similar that found by \cite{2006ApJ...640L.111T}.

Much theoretical work \citep[e.g.][and references therein]{2006MNRAS.370.1422A,2006ApJ...652L..75P,Aragon-Calvoetal2007,2007MNRAS.375..184B,2011MNRAS.413.1973W,2012MNRAS.421L.137L,2012arXiv1201.5794C,2012arXiv1201.6108T} has been dedicated to predicting, understanding and quantifying these alignments. Although these studies have access to the full six-dimensional phase space information, spin alignments with respect to the LSS, but never with the vorticity, have until now been investigated. Understanding how the Milky Way halo spins is of direct importance to studies that examine the geometry and kinematics of the Milky Way's satellite galaxy population \citep{Knebeetal2004,Libeskindetal2005,Libeskindetal2009,Libeskindetal2011a,Deasonetal2011,2012arXiv1204.6039P}.

The vorticity studied here reflects the behavior of the flow field on scales greater than $\sim4r_{\rm vir}$ which for Milky Way type haloes is approximately the Mpc scale. Just a handful of dwarf galaxies near the Milky Way have reliable proper motions \citep[see][for a description of these]{MetzKroupaLibeskind2008}: it is thus still too difficult to reliably measure the spin of the Milky Way's halo and the vorticity of our cosmic environment. However, future telescopes such as LSST and SkyMapper will be able to measure proper motions of dwarfs on these scales, thus probing the external flow field. With respect to external galaxies, missions such as GAIA and EUCLID may be able to map the large scale orientation of the vorticity in larger volumes. This can be done by mapping the orientation of disc galaxies and using it to model the orientation of the vorticity. The well known misalignment between simulated galactic disks and their parent halos \citep{2005ApJ...627..647B,2005ApJ...627L..17B,Libeskindetal2007,2010MNRAS.404.1137B} complicates the suggested task. Yet, the feasibility of such an attempt needs to be tested and gauged against cosmological simulations.

\end{document}